# LAS ACTIVIDADES DE LABORATORIO EN FÍSICA: UN CAMBIO EN LA ESTRUCTURA A FAVOR DEL CONOCIMIENTO CIENTÍFICO

**Andrea Pereira[1], Álvaro Suárez[2]**
[1]*Educación Secundaria, Montevideo, Uruguay.*
*e-mail: apefis2011@gmail.com*
[2]*Departamento de Física, Consejo de Formación en Educación, Montevideo, Uruguay.*
*e-mail: alsua@outlook.com*

**Resumen:** En este artículo se argumenta como el uso de las guías de práctico son pedagógicamente inadecuadas para un aprendizaje significativo. Posteriormente se describe un enfoque de las actividades de laboratorio acorde con el trabajo científico y con la visión constructivista del aprendizaje, ejemplificando como dicho enfoque puede aplicarse para reestructurar una práctica de laboratorio habitual.

**Palabras clave:** Didáctica, laboratorio, protocolos, problemas abiertos.

## A) Introducción

En nuestra actividad docente en distintas instituciones, hemos observado con preocupación como en los primeros cursos de física, las actividades de laboratorio suelen estar relegadas, haciéndose en muchos casos uso y abuso de las "horas pizarrón" en desmedro del trabajo experimental. Creemos que esto es consecuencia de múltiples factores, entre ellos la gran cantidad de alumnos por grupo, problemas de conducta, falta de material en el laboratorio, así como la ausencia de un preparador específico de la asignatura. Por otro lado, en bachillerato, pese a tener horas destinadas específicamente a las actividades de laboratorio, éstas en muchos casos no son explotadas como se debería, convirtiéndose en un complemento a la adquisición de conocimientos teóricos y a la resolución de problemas, presentándose como simples "recetas de cocina" a desarrollar lo más rápido posible.

En la gran mayoría de los liceos de nuestro país, está instaurado como estandarte de los cursos de práctico el uso de "guías de trabajo o protocolos". Estas guías permiten desarrollar la clase en forma ordenada y normalizar la labor de los docentes, pero, ¿son un instrumento válido para desarrollar las competencias que deben adquirir nuestros estudiantes en el laboratorio? Para contestar esta pregunta, primero debemos establecer las mismas.

A partir de los trabajos de Gil Perez (2005) y del documento sobre el rol del laboratorio en la enseñanza de la Física elaborado por la APFU (2002) se desprende que las competencias que esperamos que desarrollen nuestros estudiantes en el laboratorio son:
- Desarrollar un espíritu crítico.
- Elaborar hipótesis y modelos de interpretación.
- Contrastar y aplicar modelos.
- Reconocer los rangos de validez de los modelos elaborados.
- Predecir la evolución de fenómenos
- Diseñar y armar experimentos en función de las hipótesis y modelos que pretende verificar y del material de laboratorio.
- Controlar variables. Tomar datos.
- Interpretar los resultados obtenidos.
- Elaborar memorias.

Plantearemos en este trabajo que el uso del protocolo rígido, no sólo impide desarrollar muchas de las competencias mencionadas, siendo pedagógicamente inadecuado, sino que además genera una concepción equivocada de la ciencia. Para lograr este cometido analizaremos una guía de práctico típica.
Describiremos las propuestas desarrolladas en didáctica en las líneas de investigación actual, donde se conciben las actividades de laboratorio como pequeñas investigaciones o problemas prácticos abiertos. Plantearemos por último como pueden plasmarse estas ideas para reformular el encare de la práctica que se analizará, con el fin de demostrar que es posible reformular cualquiera de las actividades de laboratorio que se realizan en enseñanza media, para poder mejorar los logros cognitivos que pretendemos alcancen nuestros estudiantes.

## B) Algunas consideraciones sobre la forma de presentar una práctica mediante el uso de un protocolo

Consideramos que parte del desarrollo de las prácticas realizadas en los cursos de segundo y tercero de bachillerato tienen algunas características que dificultan o limitan las formas de trabajo dentro del laboratorio. Especialmente si se trata de liceos macros que presentan varios grupos por curso, en donde todo el cronograma de actividades está impuesto desde hace varios años sin modificaciones, independiente del tiempo y el espacio. No se reflexiona sobre la forma de presentar la práctica, ni se repiensa su enfoque ni su forma de trabajo, desconociendo las competencias que se deberían desarrollar.

En los programas oficiales de los cursos antes mencionados se hace hincapié en que la Física no está dividida en dos partes, el teórico y el práctico, razón por la cual el docente debe ser el mismo. Por lo tanto debe haber una coherencia entre los objetivos planteados y la forma de trabajo, coordinando lo que nos planteamos en el teórico con la parte práctica. A veces parece que se valora solo el manejo de instrumentos y toma de datos en el laboratorio, o lo bien que se puede trabajar con funciones o gráficas, dándole a las prácticas un tinte de "*estar haciendo algo*"

(Hodson, 1994:306), pero esto trae un problema: la mayoría de los estudiantes no logran conectar lo que hacen con lo que aprenden.

Las actividades del laboratorio basadas en la simple obtención de datos, no se involucran con los pensamientos de los alumnos, ni son parte de estos, son algo ajeno a ellos que solo los precisa para obtener datos y poder depurarlos de forma de obtener la relación funcional entre magnitudes pedidas por el profesor. Entonces *"el fracaso a la hora de hacer que los estudiantes participen en la reflexión que precede a una investigación experimental convierte gran parte de la práctica de laboratorio siguiente en un trabajo inútil desde el punto de vista pedagógico"* (Hodson, 1994:306). Además si no se tiene en cuenta los preconceptos que tienen los alumnos, las observaciones que realicen estarán afectadas por estos y pueden llevar a equivocaciones, concluyendo en situaciones de ensayo y error, donde tratan de llegar al resultado que procura el docente, dejando de lado la participación propia.

Siguiendo la forma general en que se plantea una práctica, queremos argumentar como en estas actividades que se realizan con los alumnos, ellos no logran adquirir las competencias o habilidades mínimas que se deben plantear los docentes de Física. Para eso señalaremos cuales son las competencias que deben adquirir los estudiantes y luego plantearemos un caso al azar de una práctica de las más clásicas e institucionalizadas que ya desde su forma de ser propuesta evidencia su falta de capacidad para lograr las metas propuestas por los docentes.

## C) Estudio de un caso que evidencia el escaso logro de habilidades adquiridas por los alumnos

Cualquier actividad que planifique un docente debe incluir las competencias que quiere desarrollar a partir de su propuesta. Por lo tanto, debería ser capaz de modificarla en función de las características del alumnado. Sin embargo, si trabaja con propuestas generales al no contextualizar los objetivos, puede generar bajos logros así como desapego y falta de interés por la asignatura.
Por otro lado existe la creencia que por proceder como se pide en cada actividad de laboratorio, los estudiantes van a desarrollar todas las competencias que se esperan. Esto es un engaño en el que caen alumnos y docentes, a su vez se transmite de generación en generación y se arraiga como parte que caracteriza el trabajo en el laboratorio.

En la introducción se hizo un punteo de las competencias que se pretende que adquieran los estudiantes en los cursos de prácticos de Física y como ya se aclaró brevemente, muchas veces estas están muy lejos de ser desarrolladas. Para estudiar este problema analizaremos un caso de una actividad práctica clásica: Aplicaciones de la Ley de Ampere.

*Caso*: Presentación de la práctica: Aplicaciones de la Ley de Ampere

*Contexto*: Se desarrolla en el último curso de Física del Liceo Miranda.

Se plantea la actividad práctica en base a un protocolo (ver anexo).

Describiremos como a partir del protocolo se pueden evidenciar la rigidez de las prácticas, realizando además pocos aportes al conocimiento científico, imposibilitando que los estudiantes desarrollen las competencias esperadas.

*Detalles*: A continuación se realiza un punteo que señala los inconvenientes de la actividad de laboratorio en forma concreta.

- El protocolo comienza con un breve resumen sobre una aplicación concreta de la ley en cuestión, a continuación se plantean los objetivos y se pide que adjunten determinada información. Aquí de entrada se plantea el tema marcando el problema a trabajar y las variables que se van a estudiar, por lo que el alumno no es capaz de contrastar y aplicar modelos, ni reconocer las magnitudes que están en juego. Queda sin sentido la posible elaboración de hipótesis ya que se establecen a priori los objetivos y la búsqueda de información.
- Se presenta un dibujo con el dispositivo de trabajo armado (ver figura 3 del protocolo) con los materiales a utilizar. Esto deja sin efecto la posibilidad de diseñar el dispositivo de trabajo, seleccionar el material necesario para trabajar las variables involucradas, o que traten de diseñar un dispositivo donde se pueda reconocer las ventajas y desventajas, según los modelos trabajados en clase o cursos anteriores.
- Relacionado con la figura antes mencionada aparece la siguiente frase: *"Representa simbólicamente el circuito de la figura y explica cómo se controla i en el mismo"* Recién aquí parece que el alumno tiene posibilidad de aplicar algún modelo simple sobre circuitos eléctricos o favorecer la capacidad de argumentar. Esto significa que recién en este tramo de la práctica puede enfrentarse a la posibilidad de desarrollar esta destreza, aunque parece difícil que lo haga porque seguramente pase por alto esta actividad por la forma en que es presentada.
- En la segunda página del protocolo (junto a la figura 4) donde muestra la relación vectorial entre los campos magnéticos, se da una breve forma de proceder para obtener datos. Aquí los estudiantes tienen la posibilidad de registrar, analizar y comparar datos. También se plantea: *"Piensa que será mejor usar, una brújula grande o una pequeña para tomar las medidas"*, sin embargo no aparece en forma de interrogante, por lo que es probable que los alumnos la tomen como una parte del procedimiento.
- Los cuadros de valores (con una columna completa) consideran sólo la posibilidad de un tipo de dato a obtener y de ahí pasan a la obtención arbitraria del campo magnético generado por el conductor (siempre todas las variables o magnitudes aparecen abreviadas).
- Por último se les solicita a los estudiantes que realicen gráficos, cálculos de pendiente y concluyan a partir de los resultados obtenidos. Solamente en este punto se les da la posibilidad que argumenten y concluyan pero de forma muy delineada.

Después de estos puntos planteados se puede pensar en que el protocolo está incompleto, o quizás demasiado armado o estructurado. También que sería mejor no utilizar este tipo de cronograma o sistema de pautas. La presentación del caso muestra lo que sucede en forma general con el desarrollo de la mayoría de las prácticas clásicas de cada curso de Física. Independientemente de la práctica, se use o no el protocolo, las actividades de laboratorio en su mayoría tienden a desarrollar siempre una pequeña parte de todas las competencias esperadas.

*¿Qué competencias generalmente no se trabajan con nuestro sistema tradicional de prácticas?*

- Desarrollar un espíritu crítico.
- Elaborar hipótesis y modelos de interpretación.
- Contrastar y aplicar modelos.
- Reconocer los rangos de validez de los modelos elaborados.
- Predecir la evolución de fenómenos
- Diseñar y armar experimentos en función de las hipótesis y modelos que pretende verificar y del material de laboratorio.

## D) Las actividades experimentales y las imágenes erróneas sobre ciencia

Cuando se trabaja en el laboratorio, no se reflexiona sobre qué tipo de concepciones de la ciencia se trasmite a los estudiantes en función de las características y enfoques de las actividades realizadas.

Según investigaciones actuales una de las claves para mejorar la educación en ciencias, pasa por cambiar la imagen que los docentes proporcionan implícita o explícitamente sobre ciencia a sus estudiantes. El dar visiones sobre ciencia que no se ajustan por ejemplo, a la forma en que evoluciona, sus conexiones con otras ramas del conocimiento o la manera en que trabajan los científicos, genera desinterés en el alumnado, el cual termina siendo un obstáculo para el aprendizaje. Esto se debe a la forma en la cual se presentan los conocimientos en el aula, completamente elaborados, sin posibilitar que los educandos realicen algún tipo de actividad característica de los científicos. Por lo tanto, los estudiantes de secundaria y por ende los estudiantes de profesorado, tienen una imagen de ciencia vinculada al "Método científico" mágico y único como receta para realizar ciencia. (Gil Perez, 2005)

Como afirma Hodson (1994:308), *"la práctica de la ciencia es una actividad poco metódica e imprevisible que exige a cada científico su propio modo de actuar. En este sentido, se puede afirmar que no hay método"*, aunque abarca cuatro elementos principales:
1. La fase de diseño y planificación donde se hacen preguntas, se formulan hipótesis y se diseñan los dispositivos experimentales.
2. La realización de los experimentos.
3. La etapa interpretativa de los resultados experimentales.
4. La fase de registro y elaboración de informes con los desarrollos, resultados

más significativos y conclusiones obtenidas con el fin de comunicarlas a otros científicos.

Estas etapas descritas no están separadas, ni existe un orden preestablecido, retroalimentándose entre sí, pudiendo por ejemplo la interpretación de los resultados experimentales llevar a una nueva fase de diseño experimental. Hacer ciencia no se caracteriza por tener un método particular y puede muchas veces llevar a caminos impredecibles a priori.

**E) Las percepciones de ciencia que generan los protocolos de práctico**

Al analizar la guía de trabajo que utilizamos como modelo para nuestra monografía, si suponemos que el docente la aplica en forma pura a su clase, encontramos que su aplicación conlleva a imágenes erróneas y distorsionadas de la ciencia.

1. Se refuerza la imagen individualista de la ciencia, ya que en la introducción y descripción de la práctica no se toman en cuenta los aportes realizados por muchos científicos, que llevaron a Ampère a enunciar la ley que lleva su nombre.
2. Se transmite una visión exclusivamente analítica de la ciencia, al no explicitar las simplificaciones realizadas en el análisis cuantitativo de la relación *B-i* y *B-d*. Por ejemplo:
    - ¿Qué tan largo debe ser el conductor para poder considerarlo "muy largo"?
    - ¿Por qué se mide el campo magnético en el centro de la brújula?
    - ¿Qué supuestos estamos realizando respecto al campo magnético terrestre?
    - ¿Por qué se realizan medidas con *i* constante y después con *d* constante?
    - ¿Tiene significado físico la constante *K*?
    - ¿Por qué es importante que la componente horizontal del campo magnético terrestre sea perpendicular al campo magnético generado por la corriente en la posición donde se coloca la brújula?
3. Se descontextualizada de la ciencia. ¿Qué relación puede tener el hecho de que la corriente eléctrica genere un campo magnético con la tecnología actual? En esta práctica se pierde la oportunidad de conectar los conceptos teóricos y muy abstractos como lo es el campo magnético, con la vida diaria de los alumnos. Cuando se hace referencia al campo magnético terrestre, no se discute que tanto sus características como las posibles causas de su existencia, son un tema interés actual en la comunidad científica. Tampoco se menciona el hecho que Uruguay es el país con menor valor del campo magnético terrestre.
4. Se refuerza una imagen empiro-inductivista y ateórica. El estudiante podría entender a partir de la guía de práctico que no es necesario realizar hipótesis alguna sobre el cuadro conductor que se utilice y que las conclusiones a las que se lleguen son independientes de las dimensiones del mismo, pudiéndose considerar siempre como muy largo.
5. Se genera una visión rígida algorítmica e infalible, ya que a partir de la guía se infiere que siempre se deben verificar las relaciones de proporcionalidad planteadas, por lo tanto en caso de no verificarlas, se debe adjudicar la discrepancia a errores en las medidas. Se sigue un método (el de la guía) para desarrollar el experimento. No se plantea la posibilidad de realizar revisiones al método experimental utilizado, ni se incentiva al estudiante que proponga nuevas mejoras.

A partir de los análisis realizados, queda claro que el trabajo con guías de práctico tal como la presentada, no es capaz de desarrollar, ni promover las competencias esperables en el alumnado, generando una imagen sobre la ciencia distorsionada y errónea, que actúa como obstáculo para el aprendizaje.

## F) Enfoque del trabajo de laboratorio como pequeñas investigaciones

Las investigaciones actuales en didáctica proponen un cambio en el desarrollo de las actividades de laboratorio, de forma tal que estén acordes con el trabajo científico y con la visión constructivista del aprendizaje. Las actividades de laboratorio se pueden armar como problemas prácticos más abiertos, que los estudiantes deben resolver sin la dirección impuesta por una guía estructurada. Se pretende que las prácticas de laboratorio se aproximen a investigaciones, integrando aspectos de las actividades científicas.

*"En una investigación abierta, al estudiante le cabe toda la solución, desde la percepción y generación del problema; su formulación en forma susceptible de investigación; la planificación del curso de sus acciones; la selección de los equipamientos y materiales, la preparación del montaje experimental, la realización de medidas y observaciones necesarias; el registro de los datos en tablas y gráficos; la interpretación de los resultados y la enumeración de las conclusiones. De esta forma se propician actividades relevantes y motivadoras para los estudiantes, que los desafíen a utilizar sus habilidades cognitivas para construir modelos más robustos, capaces de dar sentido a sus experiencias"*. (Suárez, Tornaría, 2012:2)

Este enfoque puede potenciarse con el uso de las nuevas tecnologías. La computadora en el laboratorio permite que la recolección y el procesamiento de datos se realice de forma rápida y pueda ser repetida todas las veces que sea necesario. Esto facilita el control de otras partes del proceso, como la planificación de la actividad, la selección de magnitudes a medir, la ejecución de la práctica y la interpretación de los resultados. Se dan las condiciones para que en caso de ser necesario, se modifique el plan de trabajo, tal como ocurre frecuentemente en una investigación. Hace posible también estudiar situaciones más complejas, como por ejemplo las que involucran un número grande de variables o las que ocurren muy rápidamente para ser observada por medios convencionales. (Borges, 2002).

Siguiendo estas nuevas tendencias sobre el trabajo de laboratorio, Gil Perez (2005) plantea una serie de aspectos que deben contener los mismos, en el marco de transformarlos en pequeñas investigaciones:

1. Plantear situaciones abiertas, para que los alumnos decidan como enfocar el problema hasta que lo puedan convertir en una situación manejable. Para ello los estudiantes deben plantear hipótesis, que les van a permitir guiarse en el trabajo. Estas hipótesis pueden hacer que se expliciten en muchas situaciones las preconcepciones del alumnado.
2. En el marco del acotamiento del problema, el mismo debe ser susceptible de poder realizar análisis cualitativos, donde se puedan estimar cuando corresponda, órdenes de magnitud de algunas medidas.
3. Buscar relaciones entre las situaciones propuestas y sus implicaciones con la ciencia, tecnología, sociedad y medio ambiente. También se debe integrar el tema trabajo con otras ramas de conocimiento.
4. Buscar cuando corresponda relaciones que permitan contrastar las hipótesis planteadas.
5. Los estudiantes tienen que diseñar sus propios experimentos que permitan verificar las hipótesis planteadas, así como las consecuencias que se derivan de las mismas. Deben realizar una búsqueda de trabajos de índole similar, hechos por otras personas, para contrastar los resultados obtenidos.
6. Fomentar la inclusión de nueva tecnología para el diseño de las actividades experimentales.
7. Los estudiantes tienen que reconocer la importancia del control de las variables en una actividad experimental y deben saber controlarlas.
8. Plantear que el trabajo de investigación puede modificar su rumbo en función de los resultados que se están obteniendo, incentivando la revisión de los diseños e hipótesis planteadas.
9. Realizar memorias científicas que muestren el proceso y los resultados de sus investigaciones. Estas memorias se pueden usar de base para discutir con sus compañeros sobre el desarrollo de sus trabajos.
10. Fomentar el trabajo en grupos y el intercambio entre distintos equipos.

**G) Adaptando las actividades de laboratorio**

La realización de actividades prácticas con el enfoque de pequeñas investigaciones, teniendo en cuenta los aspectos que se han planteado a lo largo de este trabajo, contempla todas las competencias que queremos desarrollar en nuestros estudiantes y están acordes con la práctica científica verdadera. Sin embargo, no se puede realizar durante un curso de un año, una actividad de laboratorio de cada uno de los temas dados en el teórico con un enfoque como el mencionado, ya que cada práctica sería muy extensa. Vemos entonces que resulta imposible que cada actividad de laboratorio se pueda trabajar plenamente como una pequeña investigación donde se desarrollen todas las competencias deseables.

Para adaptar el nuevo enfoque de trabajo a nuestra realidad en el aula, se debe realizar una cuidadosa selección de las actividades y reconocer en cada una de ellas que competencias se trabajan, de forma tal que sean complementarias. En cada práctica de laboratorio, el docente debería ser capaz de contestar la pregunta, ¿qué competencias se están trabajando?

Desde el punto de vista del estudiante, como afirma Roederer (2002:13) *"el trabajo práctico debe ser encarado de manera tal que el alumno pueda contestar sin titubeo la pregunta (verdaderamente trivial): ¿Qué he aprendido en concreto al realizar este trabajo?"*

A continuación se presentará un posible camino que se puede seguir con los estudiantes, para que realicen la práctica del conductor recto, adaptada al enfoque presentado. Se discutirá como las actividades que se irán proponiendo, permiten desarrollar una gran cantidad de competencias en los alumnos. Con esto pretendemos demostrar que se puede tomar una actividad de laboratorio clásica y sumamente lineal y reformularla presentando un enfoque alternativo de trabajo, infiriéndose que es posible repensar cada una de las actividades prácticas que se realizan comúnmente en los laboratorios.

## H) El campo del conductor recto como una pequeña investigación

Si pretendemos convertir la práctica del conductor recto en una actividad de laboratorio que tenga algunas características de una pequeña investigación, no pueden estar definidos en forma rígida todos los procesos de la práctica.

En el protocolo analizado, el objetivo inicial planteado es que el estudiante, partiendo del hecho que el campo magnético generado por el conductor recto y largo es directamente proporcional a la intensidad de corriente e inversamente proporcional a la distancia al centro del mismo, verifique que la constante de proporcionalidad es la constante magnética *K*. Independientemente de que sea este el objetivo planteado u otro, se debe desarrollar la actividad tendiendo a que sea el estudiante quien se plantee las hipótesis necesarias para poder (o no) corroborarlas.

Para plantear un enfoque alternativo al práctico del conductor recto, debemos tener en cuenta primero de que material dispone el laboratorio, ya que esto va a condicionar la manera en la cual se va a determinar el campo magnético generado por el conductor y por ende alguna de las características de la propuesta (aunque no su espíritu).Como en la guía analizada no se hace referencia a interface alguna, supondremos que el laboratorio no cuenta con una, por lo que dicho campo debe ser determinado con una brújula. Sin entrar en la discusión sobre la validez del uso de las interfaces en el trabajo de laboratorio y sobre si debe utilizarse ésta siempre que el experimento lo permita, consideramos que para esta actividad en particular puede resultar muy útil e instructivo el sensor de campo magnético si el laboratorio cuenta con uno.

En esta práctica se da por sobreentendido que el estudiante comprende plenamente por qué la componente horizontal del campo magnético terrestre tiene que ser perpendicular al campo magnético generado por la corriente en la posición donde se coloca la brújula para poder determinar dicho campo, sin embargo este hecho dista de ser obvio. Como una manera de introducir a la brújula como sensor de campo magnético, se puede comenzar la práctica con una actividad previa que lleve a que los estudiantes comprendan el rol que cumple y como puede utilizarse para determinar campos.

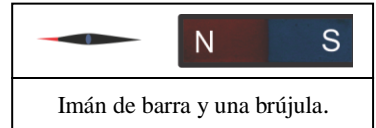
Imán de barra y una brújula.

Para comenzar se les entrega a los estudiantes un imán de barra y una brújula y se les pide que acerquen la brújula lo suficiente al polo Norte del imán, tal que el campo magnético generado por el mismo sea mucho más grande que el de la Tierra en la posición donde se encuentra la brújula.

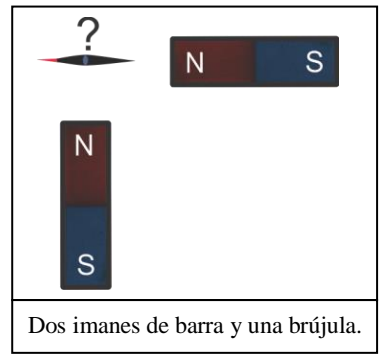
Dos imanes de barra y una brújula.

Posteriormente se les pregunta: ¿Qué pasará con la dirección en la cual estará orientada la brújula si se va acercando lentamente el polo Norte de otro imán de barra en dirección perpendicular al primero?

Con estas preguntas se espera que los estudiantes sean capaces de elaborar un simple modelo de la situación donde reconozcan que el campo magnético es una magnitud vectorial, que los campos de los imanes en el eje de la brújula son perpendiculares entre sí y que la brújula se va a orientar en la dirección del campo resultante de los dos imanes.

A continuación se les plantea el siguiente desafío: Si la componente horizontal del campo magnético terrestre vale $1,8 \times 10^{-5}$T, ¿a qué distancia del extremo del primer imán el módulo del campo magnético generado por éste es igual a la componente horizontal del campo terrestre?

Al resolver este pequeño desafío los alumnos comprenden claramente cómo pueden utilizar la brújula para determinar campos magnéticos. A continuación se les podría seguir preguntando a qué distancia el campo magnético generado por el imán vale la mitad que la componente horizontal del campo terrestre. De esta manera los estudiantes aprenden por una serie de razonamientos lógicos como usar la trigonometría para determinar el campo magnético del imán y porque razón la componente horizontal del campo magnético terrestre tiene que ser perpendicular al campo magnético que se pretende medir.

A algún estudiante se le podría ocurrir en las instancias descritas si influye o no el tamaño de la brújula. Si se diera esa situación, convendría que el profesor tuviera brújulas de distinto tamaño y planteara a los estudiantes que ellos hagan hipótesis sobre este punto y piensen como podrían corroborarlas. También podría surgir la pregunta de si los materiales magnéticos que pudiera haber alrededor afectan la medida. De surgir esta interrogante, nuevamente se les preguntaría a los alumnos como podrían corroborar esto y además se les podría plantear si el campo magnético terrestre puede variar su valor dentro o fuera del salón. Se sugiere que el profesor disponga de tornillos de distinto material para proveerles a los estudiantes.

En caso de no surgir de los estudiantes ninguna de las interrogantes planteadas, el profesor podría plantearlas o esperar al práctico del conductor recto, pero de ninguna forma puede ocurrir que el profesor conteste a la pregunta. Este debe actuar como guía y orientador y no como fuente de todas las respuestas.

Llegado este punto están dadas las condiciones básicas para plantearles a los estudiantes la práctica del conductor recto.

Cuando se realiza la práctica se da por obvio que el cuadro conductor que se utiliza se modela como infinitamente largo y simplemente se busca determinar $K$ o $\mu_0$, partiendo de la validez de la ecuación $\vec{B} = (\mu_0 i / 2\pi r)\hat{e}_\theta$. En el mejor de los casos las competencias que se desarrollan en esta actividad son la identificación de proporcionalidades y el control de variables. Tal como está hecha, el aporte de la realización de la misma a los estudiantes es poco, sino nulo.

Esta práctica tiene una potencialidad usualmente no explotada por los docentes, que es la modelización de la realidad. Muchas veces resulta difícil lograr que los estudiantes entiendan que validez puede tener determinar teóricamente el campo magnético generado por un conductor infinitamente largo o el campo eléctrico generado por un plano infinito. Esto es consecuencia de que los estudiantes no comprenden que la Física trabaja con modelos simplificados de la realidad, ni que cada análisis que se realiza de un fenómeno tiene cierto marco de validez que es función de las hipótesis en que se sustente.

La práctica del conductor recto permite ahondar sobre los modelos en física y sus marcos de validez, si se les plantea a los estudiantes la siguiente consigna: ¿Qué largo debe tener el conductor recto como mínimo para que trabajando a distancias de hasta 15cm del mismo, se pueda modelar al campo magnético como el generado por un conductor infinito?

Este problema no tiene una solución directa y los alumnos deben comparar la diferencia entre el modelo teórico y la realidad.

Con esta propuesta los estudiantes deben:
- *Diseñar y armar un experimento* para medir el campo magnético generado por el conductor a distintas distancias. Este punto permite incluir la vinculación de la práctica con temas científicos de actualidad. Al reconocer que debe utilizar la brújula y el valor del campo magnético terrestre, se les puede pedir que investiguen sobre las características peculiares del campo magnético en Uruguay[1] y su evolución con el tiempo[2]. También pueden buscar información sobre modelos actuales del campo magnético terrestre.

---

[1] En la página http://www.elpais.com.uy/100923/pnacio-517102/sociedad/uruguay-tiene-el-menor-valor-de-campo-magnetico-en-el-mundo/ se puede encontrar información sobre este tema.
[2] En la página http://www.ngdc.noaa.gov/geomag-web/#declination se pueden encontrar datos sobre el campo magnético terrestre desde el 1900 hasta la actualidad.

- *Controlar variables*. Deben reconocer que van a variar la distancia al conductor, manteniendo la intensidad de corriente constante y variar el largo del conductor para cada serie de medidas.
- *Contrastar y aplicar modelos*.
- *Elaborar un criterio* para comparar las curvas teóricas con las experimentales.
- *Interpretar los resultados obtenidos.*
- *Estimar incertidumbres*.
- *Trabajar en forma cooperativa con sus compañeros*. Como la práctica se vuelve muy larga si realizan varias medidas del campo magnético para distintos largos del conductor, esta propuesta permite la posibilidad de plantearle a distintos subgrupos de práctico y que cada uno trabaje con distintos largos del conductor. De esa manera se potencia la dimensión colectiva del trabajo científico, generando una interacción entre los distintos subgrupos.

## I) Conclusiones

A lo largo de este trabajo hemos desnudado todas las carencias que presentan las actividades basadas en protocolos y sus problemas vinculados al desarrollo de competencias en los estudiantes. Creemos que este tipo de enfoque rígido y sumamente estructurado que se le da a las prácticas de laboratorio, está arraigado a la formación docente, vinculado a la reproducción de un tecnicismo pedagógico que pasa de profesor a alumno y de adscriptor o profesor de didáctica a practicante. Muchos hemos aprendido de esa forma, con modelos de prácticas que utilizan siempre las mismas guías de laboratorio, favoreciendo solo las habilidades operatorias.

Desde nuestro punto de vista se da en las actividades prácticas demasiada importancia al registro y procesamiento de datos, descartando la construcción del conocimiento o pasándola a un segundo plano. Esto no quiere decir que las prácticas deban favorecer sólo la construcción e innovación en procedimientos e investigación. Se pretende que el docente otorgue por lo menos un grado más de libertad en sus clases y que esto comience a favorecer a los alumnos en la búsqueda del conocimiento científico que tanto promulgamos. En este trabajo hemos mostrado un camino para llegar a esto, transformando una práctica clásica de laboratorio, en una pequeña investigación, que permite un desarrollo natural de las competencias que pretendemos que adquieran nuestros estudiantes, integrando además muchos aspectos de la actividad científica.

Teniendo en cuenta a los practicantes, considerando nuestro rol de futuros formadores tendríamos que favorecer el cambio contra la reproducción sistemática. De esa manera su formación será más integral y se puede desligar de la tendencia a tener que planificar clases prácticas de manuales (ya vienen prontas) que no permiten el manejo de emergentes.

El cambio en la metodología influye en la forma de evaluar. Mientras con el sistema de prácticas estrictas los alumnos se sienten presionados, por lo cual creen necesario repetir conductas o procedimientos a la hora de terminar las actividades. En esta nueva visión de las clases prácticas la evaluación se enriquece con aportes o intercambios en todo momento, más acordes con fomentar el conocimiento científico.

## J) Referencias

**Anexo, protocolo de práctico del liceo Miranda**

# LABORATORIO DE FISICA
## LICEO Nro. 2

### PRACTICA Nro. 5 — Aplicaciones de la ley de Ampere

El campo magnético en torno a una corriente eléctrica rectilínea muy larga, puede visualizarse por las líneas de campo, usando limaduras de hierro como indica la fig.1.

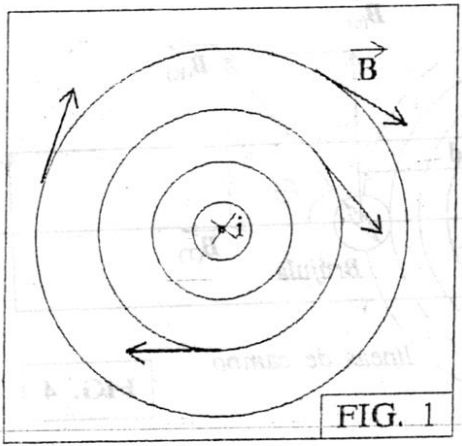

FIG. 1

La dirección del vector campo magnético en un punto cualquiera, es tangente a la línea de campo que pasa por el mismo, y el sentido se determina con la *regla de la mano derecha*.

Este campo presenta una simetría tal, que hace posible obtener su módulo en forma simple, aplicando la Ley de Ampère.

En esta actividad, verificará que el módulo de $\vec{B}$ a una distancia $d$ de la corriente está dado por:

$$|\vec{B}| = \frac{K \cdot i}{d}$$

FIG. 2

$K$ es una constante, $d$ es la distancia del punto a la corriente e $i$ es la intensidad de corriente que circula por el conductor recto y largo.

Al efectuar la verificación, determinará el valor de esa constante $K$. Para esto, usaremos un conductor recto muy largo, puesto sobre un "bastidor"; observa el dispositivo experimental y explica porque cumple con los requisitos. Piensa sobre la influencia de imanes y otros objetos, como los de hierro que se encuentren próximos al realizar las medidas. (recuerda que las mesas del laboratorio tienen "alma" de hierro).

Estudia y anexa información sobre materiales *paramagnéticos*, *diamagnéticos* y *ferromagnéticos*.

Relojes, pulseras y hebillas ¿afectarán las medidas?

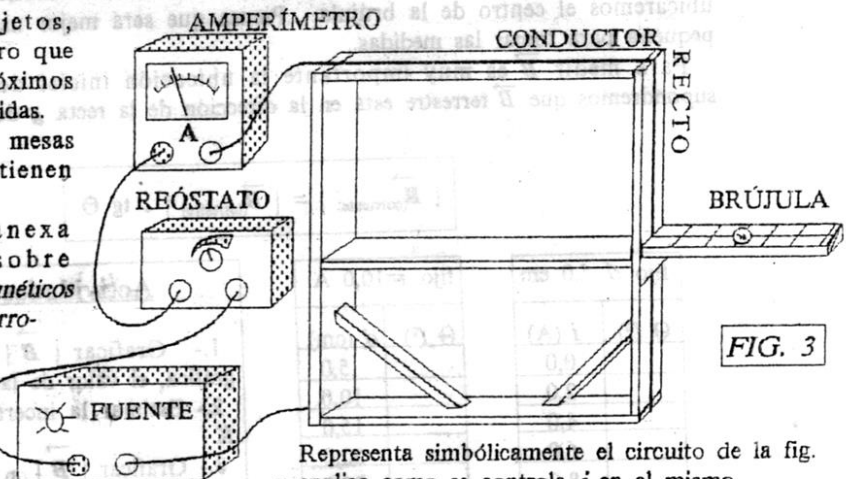

FIG. 3

Representa simbólicamente el circuito de la fig. y explica como se controla $i$ en el mismo.

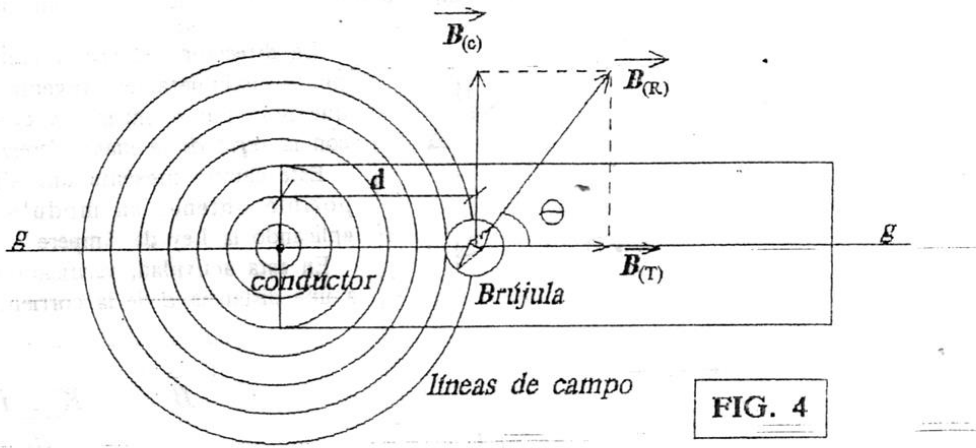

FIG. 4

En el trabajo de laboratorio, haremos dos series de medidas, la primera dejando fija la distancia $d$ y variando $i$, mediremos $|\vec{B}|$, obteniendo $|\vec{B}|_{(i)}$. En la segunda, fijaremos $i$ y obtendremos $|\vec{B}|_{(d)}$.

La medida del módulo de $\vec{B}$ se hará comparándolo con el valor del campo magnético terrestre, (cuya componente horizontal en el laboratorio será conocida), usando una brújula.

El resultado de las medidas, será para nosotros el módulo de $\vec{B}$ en el punto donde ubicaremos el centro de la brújula. Piensa que será mejor usar: una brújula grande o una pequeña para tomar las medidas.

Para medir $\vec{B}$ es muy importante la ubicación inicial del marco de madera, ya que supondremos que $\vec{B}$ terrestre está en la dirección de la recta $g$ del dibujo, obteniendo:

$$|\vec{B}_{(corriente)}| = |\vec{B}_{(terrestre)}| \cdot \mathrm{tg}\,\theta$$

fijo $d=5,0$ cm

| $\theta$ (°) | $i$ (A) |
|---|---|
|  | 0,0 |
|  | 2,0 |
|  | 4,0 |
|  | 6,0 |
|  | 8,0 |
|  | 10,0 |
|  | 12,0 |

fijo $i=10,0$ A

| $\theta$ (°) | $d$ (cm) |
|---|---|
|  | 5,0 |
|  | 10,0 |
|  | 15,0 |
|  | 20,0 |
|  | 25,0 |
|  | 30,0 |
|  | 35,0 |

### Actividades con los datos:

1.- Graficar $|\vec{B}|_{(i)}$ y obtener, usando la gráfica, el valor de la constante $K$.

2.- Estimar la incertidumbre en ese valor de $K$.

3.- Graficar $|\vec{B}|_{(d)}$ y $|\vec{B}|_{(1/d)}$ y obtener $K$ nuevamente.

4.- Concluir en función de los resultados obtenidos.

Complete los cuadros, use la ecuación anterior y calcule los módulos del campo magnético producido por la corriente en los distintos casos.

**LICEO Nro.2   H. MIRANDA**
**LAB. DE FÍSICA**